\documentstyle[12pt]{article}
\def\hybrid{\topmargin -20pt    \oddsidemargin 0pt
        \headheight 0pt \headsep 0pt
        \textwidth 6.25in       
        \textheight 9.5in       
        \marginparwidth .875in
        \parskip 5pt plus 1pt   \jot = 1.5ex}

\hybrid
\newskip\humongous \humongous=0pt plus 1000pt minus 1000pt
\def\caja{\mathsurround=0pt}
\def\eqalign#1{\,\vcenter{\openup1\jot \caja
        \ialign{\strut \hfil$\displaystyle{##}$&$
        \displaystyle{{}##}$\hfil\crcr#1\crcr}}\,}
\newif\ifdtup

\relax

\def\be{\begin{equation}}
\def\ee{\end{equation}}
\def\ba{\begin{eqnarray}}
\def\ea{\end{eqnarray}}

\begin{document}
\renewcommand{\theequation}{\thesection.\arabic{equation}}
\newcommand{\beq}{\begin{equation}}
\newcommand{\eeq}[1]{\label{#1}\end{equation}}
\newcommand{\ber}{\begin{eqnarray}}
\newcommand{\eer}[1]{\label{#1}\end{eqnarray}}
\begin{titlepage}
\begin{center}

\hfill CPTh/RR.349.0395\\
\hfill \\
\hfill \\

\vskip .5in

{\large \bf  A   way  to break supersymmetry
  }
\vskip .5in

{\bf C. Bachas} \footnotemark \\

\footnotetext{e-mail address: bachas@orphee.polytechnique.fr}

\vskip .1in

{\em Centre de Physique Th\'eorique\\
Ecole Polytechnique \\
 91128 Palaiseau, FRANCE}

\vskip .1in

\end{center}

\vskip .4in

\begin{center} {\bf ABSTRACT }
\end{center}
\begin{quotation}\noindent
I study the spontaneous breakdown of supersymmetry when
higher-dimensional Yang-Mills or
the   type-I $SO(32)$  string theory
are compactified on magnetized tori.
Because of the universal gyromagnetic ratio $g=2$, the splittings
of all multiplets are given by the product of
charge times internal helicity operators.
As a result such compactifications have two remarkable and robust
 features: {\it (a)}
they can
reconcile {\it chirality} with
{\it extended} low-energy supersymmetry in the
 limit of large tori, and
{\it (b)} they can
 trigger gauge-symmetry breaking, via Nielsen-Olesen instabilities,
 at a scale tied classically to $m_{SUSY}$.
  I exhibit a compactification of the $SO(32)$ superstring,
 in which magnetic fields break
 spontaneously $N=4$ supersymmetry, produce
the standard-model gauge group with three chiral families of quarks
and leptons, and trigger electroweak symmetry breaking.
I discuss supertrace relations
 and the ensuing ultraviolet softness.
 As with  other known mechanisms
of supersymmetry breaking, the one proposed here faces two open
problems: the threat to
 perturbative calculability in the decompactification
limit, and the
  problem of gravitational stability
and in particular of the cosmological constant.
I explain, however, why a good classical
 description of the vacuum
 may require small tadpoles for the dilaton, moduli and metric.

 \end{quotation}
\vskip1.0cm
March 1995\\
\end{titlepage}
\vfill
\eject

\setcounter{equation}{0}

{\bf 1. Introduction}

Perhaps the main puzzle
 of superstring unification \cite{GSW,Gross}
 concerns the
breaking of space-time
supersymmetry.
Two proposals have
so far been put forth in the literature:
the non-perturbative scenario based on
 gaugino condensation in a hidden sector
 \cite{Louis,Der,Potent,COND2,S,Recent},
and the tree-level Scherk-Schwarz mechanism
\cite{SS,Fayet,Rohm,Kounn,effth,ABLT,A,Karim}.
Both assume
that the correct vacuum of the theory
is an {\it exact solution} of the classical
 string equations, which
  leave   undetermined the values of
various continuous moduli. These
  should then hopefully be fixed by radiative
or non-perturbative corrections.
The mechanism of Scherk and Schwarz
  allows in particular for a
  a string realization of the no-scale supergravity
models \cite{Nano,Lah}.
Supersymmetry breaks classically at a scale  proportional to the
  values of one or more moduli, say ${\rm Re}T$, and the
  gauge hierarchy is presumably attributed to
 the logarithmic running
of couplings which yields a very shallow deformation of
the classically-flat potential of ${\rm Re}T$.
 A merit of this scenario
is that it is calculable at tree- and one-loop
order, at which point
one encounters instabilities
 due among other things to a non-vanishing cosmological
constant. Furthermore the embedding in
the string, which makes one-loop
gravitational corrections finite, raises a novel problem:
 within the one-loop approximation there is no hope to stabilize
the dilaton
 and avoid among other things conflict with
 the   principle of equivalence
\cite{Damour}.
 Gaugino condensation
or other non-perturbative phenomena are
 conceptually more promising
in this respect,
but we lack the technology to study them directly
at the string level.
 Most of the interesting attempts are therefore limited to
guessing   superpotential modifications, using as a guide
  field theory as well as   the $S$-duality
conjecture \cite{S,Recent}.
Even with such guesswork a realistic vacuum without
runaway dilaton and moduli
and with vanishing cosmological constant has not yet been found.
It is furthermore questionnable whether anything short of
solving {\it all} of the above
 gravitational instabilities at once, would constitute
real progress.

In view of this unsatisfactory situation,
 it is I believe fair to ask
whether we have not taken too seriously
 the classical string equations
of motion.
 What if  the true string vacuum leads to small classical
tadpoles which should be cancelled
 ultimately by higher-loop and
non-perturbative corrections?
The Coleman-Weinberg mechanism \cite{CW}
 serves
 to illustrate forcefully this point:
one considers a complex
scalar field with quartic potential,
 $V_{tree}(\phi) = \lambda (\phi^*\phi)^2$,
 coupled to a $U(1)$ gauge field
as well as to massless fermions through Yukawa couplings.
The classical field equations tell
 us that $<\phi>=0$, so that the
photon and all fermions stay massless.
 One-loop corrections on the other hand
 can  change the shape of the potential
   leading
to a non-zero vacuum expectation value for $\phi$.
Expanding around this non-zero vev
gives thus a much more accurate description of the spectrum,
  even though
classically there remains an uncancelled tadpole.
{\it Likewise, the price for getting
 a good description of the
low-energy world from the string
 may be to allow for   small  metric, dilaton and
moduli tadpoles in the classical description of the
 ground state}.
To see how small is small, suppose that
in what concerns the gauge sector,
the  vacuum of the heterotic string could be approximated well by
compactification on a four-torus times a two sphere with
a magnetic monopole in its middle \cite{horv}.
 These backgrounds
  correspond \cite{RW,Monopole} to a  $SU(2)_k$
$WZW$ model and free fields, so that
apart from the   $\beta$-function of the dilaton:
$\beta_{\Phi} =
 ({\hat c}-10)/{\alpha^\prime} \simeq 4/k\alpha^\prime\not= 0$,
all other classical equations are satisfied \cite{beta}.
Now we imagine that quantum-gravity effects   cancel the
tadpoles of the dilaton and metric,
 but do not affect the gauge sector
of the theory where supersymmetry is broken at a scale
 $m_{SUSY}  \sim 1/\sqrt{k\alpha^\prime}$.
 For this scale to be $\sim 1 TeV$
we need  $k \sim 10^{30}$.
 The classical prediction of  string theory,
that the effective space-time dimension
${\hat c}=10$, would in this hypothetical case
 be a very good approximation of reality.
But given our lack of control over the full quantum dynamics
why should we expect it to be more accurate
 than one part in ten to the thirty?

Following this logic opens up a host
 of possibilities: any background
deviating a little from
a supersymmetric, classical solution of string
theory could  a priori be a good starting
 point to describe the vacuum.
This nearby
  supersymmetric solution (NSS)
 must of course contain the right gross
ingredients: a gauge group
$G \supseteq SU(3) \times SU(2)\times U(1)$   and
enough chiral fermions to describe
 the families of quarks and leptons.
The number of non-compact dimensions in the NSS may or may not
be equal to four, since the decompactification theorems
 \cite{Banks,ABLT} only apply when one restores supersymmetry
along marginal directions. As for all other  low-energy
phenomena, such as electroweak and supersymmetry
 breaking and the generation of mass, these
are fine structure when viewed from
 the Planck scale and they therefore
 depend crucially on what
kind of deviations one is willing to consider.
To make some progress we  need an ansatz for
 these deviations and I will here
   make the   choice of
turning on constant magnetic fields
in torroidally-compactified
dimensions. Though by no means compelling, this choice
  has the following attractive features:

  {\it (a)} It corresponds, as I will explain below,
to  classically-stable solutions of higher-dimensional
gauge theory, as well as of type-I
open string theory, if one arranges for   Nielsen-Olesen
 instabilities \cite{olesen}
to be absent. Tadpoles only arise upon
  coupling
the dilaton and metric,
which makes it more
 plausible that   after all the
quantum-gravity dust has settled, the classical spectrum
is still a good approximation of reality.

 {\it (b)}
 The pattern of supersymmetry breaking is elegant and
simple, with  all splittings being
  proportional to charge times
internal-helicity operators.
 This is related to the fact that
  consistently-coupled relativistic particles must have
 a gyromagnetic ratio $g=2$ \cite{Telegdi,gyro}, and it implies
powerful supertrace relations.

{\it (c)} Such compactifications can reconcile {\it  chirality} with
  {\it extended} low-energy
supersymmetry in the limit of large tori.
 This is a consequence
of the index theorem \cite{chir,GSW}, or put differently of the fact
that the magnetic field shifts the masses of mirror fermions
  in opposite directions.

{\it (d)} The Nielsen-Olesen instability
 triggers gauge symmetry breaking
when the magnetic field is embedded in non-abelian group factors.
This provides a new mechanism to break electroweak symmetry
and tie up classically $m_Z$ to $m_{SUSY}$.

  The use of open strings facilitates the analysis
and postpones gravitational
tadpoles   to the one-loop (annulus or
M\"obius strip) level \cite{annulus}.
 The above features should,
however, continue to hold if one compactifies the heterotic string
on magnetized spheres rather than tori.
 Furthermore, I expect the last two
features to be robust and to characterize a much wider class of
compactifications.
  Note that in contrast to the above mechanism, the
breaking \`a la Scherk-Schwarz is explicit rather than spontaneous
at the global level, it can
accomodate
chiral fermions
 only for minimal   low-energy
 supersymmetry   ( $N=1$ in four dimensions) \cite{Kounn}
, and it cannot trigger
electroweak breaking  classically. Both mechanisms face, however,
the same two serious difficulties:  the problem of gravitational
stability due in particular to
a non-zero vacuum energy
$ V \sim m_{SUSY}^{\ \ \ 4}  \ \ $,
and the problem of large internal dimensions \cite{Banks,ABLT}
which,
 though not in obvious conflict with
 experiment \cite{Fayet,Karim},  threaten
 weak-coupling calculability.
This latter difficulty is due in our case
to the Dirac quantization condition, which ties up the
size of   magnetic fields to the inverse area of tori.
For chiral theories it poses, as we will
 see, a problem at one-loop
order, while in the Scherk-Schwarz scenario this problem can be
postponed to two loops \cite{A}. Likewise the vacuum energy
has in our case a classical contribution,
 while   the Scherk-Schwarz
breaking generates it only at one-loop order.
The two mechanisms do not differ qualitatively in these respects,
and any way out of these difficulties
 could a priori apply to either
one or to both.
 A couple of points are nevertheless worth
stressing: first, reconciling chirality with extended low-energy
supersymmetry raises the exciting possibility that the tractable
dynamics of $N=4$ and $N=2$ supersymmetric gauge theories
\cite{SW} could be of help in addressing these issues.
Second, the classical energy of the magnetic
 fields, which is a drastic
departure from the no-scale idea,  changes the nature of the
gauge-hierarchy puzzle
\footnote{Whether it can also help with the
 problem of the runaway dilaton
\cite{DS} is unclear. Although this energy is
 multiplied  by inverse
powers of the coupling when one works with
 $\sigma$-model backgrounds,
this changes if one rescales fields
 so as to normalize the Einstein term
in the action \cite{beta}. Note also
 that in WZW compactifications of the
heterotic string the sign of this classical energy changes sign.
}
:  rather than explain why
  $m_{SUSY}$
 is so much smaller than $M_{Planck}$
we only have to explain why it is not {\it exactly zero}.
Such a tree-level energy
 was in fact added by hand in recent phenomenological
attempts to stabilize the mass of the gravitino \cite{Ilarion}.

Finally a note on references: there
 is of course a huge literature
  on compactifications of higher-dimensional gauge theories,
Kaluza-Klein theories and   supergravities
in the presence of gauge-field backgrounds \cite{KK},
and in particular  on the two-sphere
 with a magnetic monopole in its middle
\cite{horv}.
The  role  of gauge backgrounds
 in creating chirality through the index
theorem is also a well-established fact of life \cite{chir,KK,GSW},
while monopole compactifications of the $SO(32)$ superstring were
considered early on by Witten \cite{O32}.
 What was perhaps  not appreciated
in these  earlier studies,
was that magnetic fields can provide
an elegant   mechanism for spontaneous
 breaking of both supersymmetry
and electroweak  symmetry.
These  efforts
were in any case silenced by the advent of string theory,
as emphasis shifted to zero-energy exact  solutions of the
classical   equations of motion.
The possibility of violating these equations was evoked by
Rohm and Witten \cite{RW}.
 Their study of antisymmetric-tensor-field
backgrounds in relation to gaugino condensation
 in the heterotic string
is closest in spirit to the present paper.

The plan of the paper is as follows:
 in section 2 I review some elementary
facts about magnetic monopoles on the torus, and describe the
splitting of $N=2$ hypermultiplets. In order to generalize the mass
formula to vector and higher-spin massive multiplets, it is
technically more convenient to pass to the open superstring. This
is explained in section 3, which contains some known material for
completeness. In section 4 I discuss the Nielsen-Olesen instability,
anomaly cancellation,
 as well as supertrace relations
 and the ensuing one-loop ultraviolet
softness. I explain why despite this soft behaviour, and the absence
of renormalization when the tori are demagnetized, the question of
perturbative calculability in the decompactification limit
remains open.
In section 5 I exhibit a
compactification of the $SO(32)$
superstring with a three-chiral-family
standard model in the massless
spectrum, broken $N=4$ supersymmetry and a negative mass square for
the higgs doublets. Though this model has too much structure at the
supersymmetry threshold to be considered at this point as
phenomenologically viable, it
is intriguing how close it comes to
 describing low-energy physics.
Finally in section 6 I give some concluding remarks.

\vskip 0.6cm

{\bf 2. Breaking SUSY with the magnetized torus}

The simplest setting of interest is that
 of a six-dimensional gauge theory
($ A_\mu\ \ {\rm with}\  \mu = 0,1 ...,5$) compactified down to
four dimensions on a two-torus. We assume for simplicity that
the torus is generated by orthogonal vectors on the plane:
$x_4= x_4 + R_4$ and $x_5=x_5 + R_5$.
 Vacuum solutions, invariant under
 the four-dimensional Poincar\'e
group, can be characterized by a
 constant magnetic field  $F_{45} = H$,
and by the Wilson phases around a basis of non-contractible
loops on the torus. Choosing  a particular gauge we may write:
$$ A_4(x) = a_4 \ \ {\rm and}\ \  A_5(x) = a_5+ Hx_4 \ ,
 \eqno(2.1)$$
where $a_4$ and $a_5$ are constant.
Charged  quantum mechanical particles
behave very differently according
to whether $H=0$ or $H\not= 0$. In the
 former case their wavefunction
is periodic, leading to a lattice of
covariant momenta shifted from the origin by the
charge ($q$) times the constant gauge field.
Thus the   spectrum of the two-dimensional laplacian,
which gives the mass shifts of towers of Kaluza-Klein states,
reads:
$$ \delta {\cal M}^2 = {\bf p}_4^{\ 2} + {\bf p}_5^{\ 2} =
  \Bigl( {2\pi n_4\over R_4} - q a_4\Bigr)^2 +
       \Bigl( {2\pi n_5\over R_5} - q a_5\Bigr)^2 \ . \eqno(2.2)$$
Here $\delta {\cal M}^2$ stands for
 the mass of the $(n_4,n_5)$ excitation
minus the mass of the parent field in six dimensions.
This spectrum changes continuously with the values of the (periodic)
  moduli $a_4$ and $a_5$, which
parametrize inequivalent vacua. For $a_\mu \not= 0$,
all charged excitations are massive,
 including any charged (non-abelian)
gauge bosons. This is the well-known mechanism of gauge-symmetry
breaking by Wilson lines. It respects all
supersymmetries of the parent theory
since the above mass formula is spin-blind.

The situation changes drastically for a non-zero value of $H$:
the continuum of inequivalent vacua
 is replaced by a discrete set of
states, and in the presence
of charged fields all supersymmetries are   spontaneously broken.
To see why let us recall some elementary facts \cite{Manton}
 about the
gauge field, eq. (2.1). This is the field of a monopole, i.e. a
non-trivial $U(1)$ bundle over the
 torus with a transition function
  joining
the $x_4\simeq 0$ to the $x_4\simeq R_4$ regions:
 $$ A_\mu\vert_{x_4+R_4} = A_\mu -
i e^{-i\theta} \partial_\mu e^{i\theta}
\vert_{x_4} , \ \ {\rm with  } \ \ \theta=  HR_4x_5 \ .
\eqno(2.3)$$
Demanding
 that $e^{i\theta}$ be single-valued
 on the $x_5$ circle leads to the
Dirac quantization condition:
$$  H = {2\pi K \over   R_4 R_5} \ \ , \ \ \ \
 \ {\rm with }\ \   K \ \ {\rm  integer} \eqno(2.4)$$
if the unit of charge is set   equal to
 one.  Thus for fixed area of the torus,
$H$ is a discrete modulus rather than   a continuous parameter of
compactification.
  Furthermore the spectra of Kaluza-Klein excitations depend only
on the commutator
of  covariant momenta,
$$[{\bf p}_4,{\bf p}_5] = iqH \ , \eqno(2.5)$$
 but not on $a_4$ and $a_5$. Indeed $a_5$ can be absorbed by a shift of
$x_4$, and $a_4$ by a change of gauge \cite{Manton}.
Thus, although from the $4d$ point of
 view, $\delta A_4$ and $\delta A_5$
are scalar fields with a flat potential, their expectation values
are physically irrelevant and do not label inequivalent vacuum states
 This
 is of course a common phenomenon:
the Goldstone boson of a spontaneously-broken
 global symmetry
has also a physically irrelevant expectation value,
 as does the dilaton
in a linear-dilaton vacuum of string
 theory.

That the magnetic field breaks supersymmetry could be argued for on
the basis of its positive contribution to vacuum energy. Strictly
speaking this is incorrect, since in the absence of charged fields
(and of gravity) all of the $H$-vacua
would still be supersymmetric
\footnote{This is of course possible because $6d$
Poincar\'e invariance has been broken.}.
Let us assume therefore that the parent $N=1$ supersymmetric $6d$
theory contains some charged chiral (and hence massless)
 hypermultiplet.
This yields by trivial reduction
  two complex scalars
and two Weyl spinors of opposite   chirality  in four dimensions
\cite{Fayet}.
 Compactifying in the   background of the
magnetic field
splits this hypermultiplet in an interesting way. First,
the Laplace operator on the torus has
  now the spectrum of a harmonic
oscillator, giving the following mass shifts
 for the scalar components
of the multiplet:
$$ \delta{\cal M}_{(0)}^{\ 2 }=
{\bf p}_4^2 + {\bf p}_5^2 =
 (2n+1)\vert q H \vert \  \ \ {\rm with}\ \ n=0,1,2 ..  \eqno(2.6)
$$
 Each Landau level   in the above spectrum  is, furthermore,
 $(qK)$ times degenerate,  with both  $q$ and $K$ being integers
by virtue of quantization of charge.
 A set of wavefunctions that span, for example,
 the lowest Landau level
when $a_4=a_5=0$ are the following:
$$
 \Phi_{(0),j} =  {\cal N} \ \sum_{m=-\infty}^{\infty} \
 exp\Biggl[ {-{1\over 2} \vert qH \vert
\Bigl( x_4 - (m + {j\over qK}) R_4\Bigr)^2}\Biggr] \ exp\Biggl[
 {2\pi i (qKm +j) {x_5\over R_5}}\Biggr] \   \  ,\eqno(2.7)
$$
where ${\cal N}$ is a normalization and
 $j= 1,2 .. ,qK$.
These   have  indeed the required periodicities:
$\Phi(x_4+R_4,x_5) = {\rm exp}(i q \theta) \Phi(x_4,x_5)$ and
$\Phi(x_4,x_5+R_5) =   \Phi(x_4, x_5)$.
For higher Landau levels one must replace the first
of the two exponentials above with higher excited
harmonic-oscillator eigenfunctions.

Consider next what happens to
the $6d$ Weyl spinor. Denoting by $\Gamma_\mu$ the
  Dirac matrices which obey
$\{\Gamma_\mu,\Gamma_\nu\} = 2\eta_{\mu\nu}$,
and using the commutator (2.5), one finds:
$$\eqalign{  \delta {\cal M}_{(1/2)}^2 =
({\bf p}_4  \Gamma_4 + &{\bf p}_5 \Gamma_5)^2
   \cr  &=
  (2n+1)\vert qH \vert + 2qH \Sigma_{45} \ \ ,\cr } \eqno(2.8)$$
Each Landau level is again $(qK)$ times degenerate,
and
 $\Sigma_{45} = {i\over 4} [\Gamma_4,\Gamma_5]$
 is the $6d$ spin operator projected along the magnetic
field.
Since the six-dimensional spinor was Weyl,
 we can identify the internal
helicity $i\Gamma_4\Gamma_5$   with the
four-dimensional chirality, so that
  $\Sigma_{45}=\pm {1\over 2}$ for chiral
 or antichiral $4d$  spinors.
At the lowest Landau level we thus obtain $(qK)$ massless chiral
fermions, while their antichiral partners are shifted to  a
mass equal to $2 \vert qH \vert$. There they are joined by the
chiral fermions of the first Landau level so as to form
$(qK)$  massive Dirac spinors, and the tower continues like that
forever. All this is of course in agreement with the
(two-dimensional) index
theorem:
 $$
 index ( \partial_{\bf A} ) =
{q\over 2\pi} \int dx^4 dx^5 F_{45} \ ,
\eqno(2.9)$$
which we could have used to
predict   the net number of massless chiral fermions
surviving compactification on the magnetized torus \cite{chir,KK,GSW}.

It follows easily from the above expressions
 that for each Landau level
separately
$$ Str\ {\cal M}^2 = 0 \ , \eqno(2.10)$$
  in accordance with the fact that the breaking of supersymmetry
is spontaneous. Note that, in contrast, the Scherk-Schwarz
mechanism breaks global supersymmetry explicitly, by
modifying the boundary conditions of fields
  as in the case of
finite temperature,
so that the above supertrace may \cite{Fayet} but need not vanish.
Note also how chirality can be reconciled,
 as advertized, with low-energy
$N=2$ supersymmetry in four dimensions.
In the stringy
  Scherk-Schwarz scenario by contrast, chiral fermions live
in twisted sectors of orbifolds, which are
 spectators of the symmetry-breaking process
\cite{Kounn}.  Equations (2.8) and (2.10) will stay   valid for
vector as well as massive higher-spin
  multiplets. This can be shown
easier in the context of
  the open superstring to which we now turn our attention.

\vskip 0.6cm
{\bf 3. Type-I superstring.}

A constant electromagnetic background adds only
quadratic boundary  terms
to the   world-sheet action of an open string, so that
 the corresponding conformal field
theory can be solved exactly \cite{Tseytlin,Callan}.
This has for instance  been exploited to
  calculate   the rate of string-pair creation in
a constant electric field \cite{Schwinger}, and
 to  show that the gyromagnetic ratio  is  $g=2$
 for  all higher-spin string excitations  \cite{Telegdi}.
The latter is the key observation which we want now to adapt
in our context.  We consider for definiteness
some torroidal compactification of
  the $SO(32)$ superstring
  from ten down to four dimensions,
and turn on for the time being a
   magnetic field in only one of the three planes of the
hypertorus: $H=F_{45}$
 \footnote{ The magnetized torus
was also considered in   ref.
\cite{Callan}. This study
was restricted to the bosonic string, so that the issue
of supersymmetry breaking did not arise.}.
For later convenience we also add some
 Wilson-line breaking of the
gauge group, by exploiting all six compact dimensions:
$a_I$ for $I=4,..,9$.
We take all these backgrounds   in the Cartan
subalgebra of $SO(32)$, and     denote by
$q_{L(R)}$
 the left(right) end-point charges of the string in the
  direction of $H$, and by $q^{I}_{L(R)}$
the corresponding charges in the
direction of $a_I$.
The
world-sheet action on the   strip reads:
 $$\eqalign{
S_{world-sheet} = -{1\over 4\pi\alpha'}& \int
   d\tau \int_{0}^{\pi}
 d\sigma
\bigl\{ \partial_a X^{\mu}
\partial^a X_{\mu} -
 i{\bar\psi}^{\mu}\rho^{\alpha}\partial_{\alpha}\psi_{\mu} \bigr\}
\cr  - &    \int d\tau \  \Biggl(\  q_L H
\bigr\{ X^{4}
\partial_{\tau}X^{5}
- {i\over 2} {\bar\psi}^{4}\rho^{0}\psi^{5}
\bigr\} + \sum_{I=4}^{9}
 q_L^I a_I \partial_{\tau} X^I \Biggr)_{\sigma=0}
\cr
-\ &     \int d\tau\  \Biggl(\  q_R H \bigr\{
X^{4} \partial_{\tau}X^{5}
-{i\over 2}  {\bar\psi}^{4}\rho^{0}\psi^{5}
\bigr\} + \sum_{I=4}^{9}   q_R^I a_I
 \partial_{\tau} X^I \Biggr)_{\sigma=\pi}
. \cr}  \eqno(3.1)$$
Here   $\alpha^{\prime}$ is the Regge slope which we
set equal to ${1\over 2}$,
  the
  $\psi^{\mu}$ are real Majorana fermions,  and
  ($\rho^0,\rho^1$)  are the corresponding
Dirac matrices in two dimensions with the conventions
of ref. \cite{GSW}.

Since the gauge fields only couple at the boundary, all
fermionic and bosonic coordinates satisfy free-wave equations.
For $I=6,7,8,9$, the $X^I$ and $\psi^I$ have the
usual Neumann, and
Ramond (R) or Neveu-Schwarz (NS) boundary conditions. The
corresponding
  momenta $p^I$
lie on a lattice which is shifted by an amount
 $(  q_L^I+   q_R^I)a_I$
from the origin, so that all charged states are
generically massive.  This is up to here a conventional torus
compactification with Wilson-line breaking of the gauge group
and maximal $N=2$ supersymmetry in six dimensions.
We may also reduce the $6d$ supersymmetry to $N=1$, by
orbifolding these extra compact dimensions \cite{HarveySagn,Dixon},
without affecting the discussion that follows.

In contrast
 to the Wilson-line backgrounds, a non-vanishing magnetic field
changes the boundary conditions of the remaining
  complex compact coordinate
$X \equiv { 1\over \sqrt{2}}( X_4 + i X_5) $  and of its
superpartner $\psi \equiv {1\over \sqrt{2}} ( \psi_4+i\psi_5)$.
 Recall
that variations of the latter must be constrained at the
boundary as follows \cite{GSW}:
$
\delta\psi_R  = \delta\psi_L $ at $ \sigma=0$, and
$\delta\psi_R = -(-)^{a} \delta\psi_L$ at $ \sigma=\pi$, where
  $\psi_{L(R)} $is
the left(right)-moving component of the   fermion
and $a=0$ or $1$
  in the Neveu-Schwarz or Ramond sector. Extremizing the action
leads therefore to the equations \cite{Callan,Telegdi,Schwinger}:
 $$\eqalign{
 \partial_\sigma X & = -i\  \beta_L\ \partial_{\tau} X \
 \  {\rm and}
\ \ \psi_R = {1+i\beta_L\over 1-i\beta_L} \ \psi_L \ \
 \ \  {\rm at} \ \ \sigma  = 0\   \ ;  \cr
   \partial_\sigma X &= i\ \beta_R\ \partial_{\tau} X \
 \  \ {\rm and}
\ \  \psi_R = -(-)^a\  {1-i\beta_R\over 1+i\beta_R} \ \psi_L\ \
 \ \  {\rm at} \ \ \sigma  = \pi\   \ ,
 \cr} \eqno(3.2) $$
where we have used the short-hand notation:
$$\beta_{L(R)} =  \pi \  q_{L(R)} H \ \ . \eqno(3.3)
$$
Since the boundary conditions are linear, we can   expand
the   coordinates in
orthonormal modes as usual:
$$ X  =
x  +   i
\sum_{n=1}^{\infty}   a_n  \phi_n (\sigma,\tau)
-i \sum_{n=0}^{\infty}   {\tilde a}_n^{\dag}
 \phi_{-n} (\sigma,\tau)  \eqno(3.4a)$$
with
$$ \phi_n (\sigma,\tau) = (n-\epsilon)^{-{1\over 2}}\  e^{-i(n-
 \epsilon)\tau} \ cos[(n-\epsilon)\sigma +
\ arctan(\beta_L)]  \ , \eqno(3.4b)$$
and
$$
\psi  = \cases{ \ &
$\sum_{n=1}^{\infty} b_n \psi_n(\sigma,\tau) \  +
           \sum_{n=0}^{\infty} {\tilde b}_n^{\dag}
\psi_{-n}(\sigma,\tau)$ \   \ \
  (Ramond)  \cr
\ &\  \cr
\ &  $\sum_{s={1\over 2}}^{\infty}\ \Bigl[
 b_s \psi_s(\sigma,\tau) \  +
           {\tilde b}_s^{\dag}
\psi_{-s}(\sigma,\tau) \Bigr] $ \  \ \  \ \  (NS)
  \cr}
 \eqno(3.5a)$$
with
$$
\psi_{({R\atop L}),{n}}(\sigma,\tau)
 = {1\over \sqrt{2}} exp \Bigl[ {-i(n-\epsilon)(\tau \mp \sigma)}
    \pm  i\
arctan \beta_L \Bigr]
  \ . \eqno(3.5b)$$
The above expressions depend on the magnetic field
   through the non-linear
function \cite{Callan}
$$ \epsilon = {1\over \pi} \ [arctan(\beta_L) +
 arctan(\beta_R)] ,\eqno(3.6)
$$
which summarizes the effects of the non-minimal string coupling.
Note that in the weak-field limit that interests us
in this paper ($H\sim m_{SUSY}^{\ \ 2} \sim 10^{-30}$) we have
 $$\epsilon = (q_L+q_R)H + o(H^3) \ , \eqno(3.7)$$
while for a field of the order of the string tension $\epsilon$
 saturates to
the values $\pm 1$ .
Canonical quantization
leads to the commutation relations:
$$ [a_n, a^{\dag}_m] = [{\tilde a}_n,
{\tilde a}^{\dag}_m]= \delta_{nm}=
  \{ b_n, b_m^{\dag} \} =
\{ {\tilde b}_n, {\tilde b}_m^{\dag} \}= \delta_{nm} \eqno(3.8)$$
and
$$ [x , x^{\dag}] = {1 \over
(q_L+q_R)H}\  .  \eqno(3.9)$$
All other commutators are zero.

The upshot of this tedious algebra is that
the complex supercoordinate $(X,\psi)$ behaves like the coordinate
of an orbifold \cite{Dixon} in a twisted
 sector with twist angle $\epsilon$.
 There are, to be sure, some significant differences between the
magnetized torus  and an orbifold:
the center-of-mass position has in our case a non-trivial commutator,
 $\epsilon$   is not related to a  discrete
symmetry and can be arbitrarily small for a large torus, and
 we do not sum over twisted sectors.
The orbifold analogy is nevertheless useful
in  deriving
the    spectrum of masses. To this end we note the
following:

{\it (i)} The creation operators
${\tilde a}_n^{\dag}$ and ${\tilde b}_n^{\dag}$ ($n>0$)
  raise the
helicity in the $(X_4,X_5)$ plane by one unit and have  their
world-sheet frequencies
shifted by $+\epsilon$. Similarly
 $a_n^{\dag}$ and $b_n^{\dag}$ lower the
helicity by one unit and have their frequencies
 shifted by $-\epsilon$.

{\it (ii)}
The zero modes require special treatment:
 if $\epsilon $ is positive,
${\tilde a}_0$ annihilates the vacuum while
${\tilde a}_0^{\dag}$ creates the successive excited Landau levels.
Likewise   ${\tilde b}_0$ annihilates the lowest-lying
Ramond state of internal helicity $-{1\over 2}$, while the action of
${\tilde b}_0^{\dag}$ flips the helicity to $+{1\over 2}$
 and raises the square mass of the state
by $2\epsilon$ \footnote{With our conventions the square mass is
given by twice the zeroth-moment Virasoro generator.}.
If $\epsilon$ is negative one must reverse the roles of daggered and
undaggered operators in the zero-mode sector.

{\it (iii)}
By virtue of world-sheet supersymmetry the magnetic
field does not shift the position of the vacuum in the Ramond sector,
while a straightforward calculation
 \cite{Dixon,Callan,Schwinger} gives a shift
of $\vert\epsilon\vert$ for the square mass
 of the vacuum in the  Neveu-Schwarz
sector.

Putting these observations together
 we arrive at the following  remarquably simple
expression for the mass shift of all string excitations when going
down from six to four dimensions:
 $$
 \delta{\cal M}_{string}^{\ 2} =
 (2n+1)\vert \epsilon \vert + 2\epsilon \Sigma_{45}\ \ ,
  \eqno(3.10)$$
with $\Sigma_{45}$   the spin operator projected on the plane of the
torus. The non-trivial commutator
(3.9) ensures
 furthermore that
each Landau level is degenerate $(q_L+q_R)K$ times.
The above expression reduces to eqs. (2.6)
 and (2.8) in the weak-field
limit and for $6d$ scalar or, respectively,  spinor excitations,
but generalizes these results to vector and higher-spin
multiplets.

Several remarks are in order here: first eq. (3.10) is
  of course only valid for $q_L+q_R\not= 0$,
in which case the Wilson lines $a_4,a_5$ are
 irrelevant. For strings
neutral with respect to the magnetic $U(1)$
  these Wilson
lines shift the lattice of $(p_4,p_5)$ momenta as previously
described.
Secondly,  the above analysis can be extended trivially
to the case where all three tori are magnetized. Labelling these
tori by a lower-case Latin index, and working from now on
in the weak-field limit, eq. (3.7),
 we may write:
$$
\delta{\cal M}_{string}^{\ 2} \simeq
\sum_{a=(45),(67),(89)} \ (2n_a+1) \vert q_a H_a \vert
+ 2 q_a H_a \Sigma_a \ \ .\eqno(3.11)$$
Here $H_a$ are the magnitudes of the three magnetic fields pointing
in some directions
  inside the Cartan subalgebra of $SO(32)$,
and $q_a$ are the corresponding total charges.
One can consider
 more general situations, such as magnetic fields
not aligned with the planes of the tori, but we wont need these
in the discussion that follows.
When all the $q_a\not=0$, a
  $10d$ Weyl spinor is split by the three magnetic fields in such
a way that only one $4d$ chiral fermion
 remains massless. Indeed, one must fix all three internal helicities:
$\Sigma_a = -{1\over 2} sgn(q_a)$, so as to cancel
the positive mass shift common to all excitations at the
lowest Landau level.
This shows how chirality
can be reconciled with {\it maximal} ($N=4$)  low-energy supersymmetry
in the limit of three large tori.
Note finally that the tracelessness of $\Sigma_{\mu\nu}$ ensures that
$str {\cal M}^2 = 0$ for any multiplet and every Landau level.

\vskip 0.6cm
{\bf 4.  Nielsen-Olesen  instability, anomalies and UV softness}

Let us take now a closer look at vector multiplets.
A charged $6d$ gauge boson (and its conjugate) gives by trivial
dimensional reduction two complex scalars
with internal helicities $\Sigma_{45} = \pm1$, and a $4d$
gauge boson with $\Sigma_{45} = 0$.
The mass formula,
 eq. (3.10) implies that the lowest Landau excitation
of one of the two scalars
has a negative mass shift equal to
$-\vert q H\vert$.
  If the gauge boson was originally
massless, the vacuum would therefore be unstable,
 as Nielsen and Olesen
were the first to point out \cite{olesen} \footnote{
Since we   work with weak magnetic fields we wont worry
about the extra instabilities which can occur when $H$ is of the
order of $\alpha^\prime$ \cite{FP}.}.
We can of course eliminate the instability by rendering the charged
$6d$ gauge bosons massive. This can be achieved
by turning on Wilson lines
in the extra compact dimensions, so that the unbroken $6d$ gauge
group is of the form $ U(1)_{H}\times G$.
 Turning on
several magnetic fields will also eliminate some of the
instabilities, since only one internal helicity can be non-zero
for low-lying $4d$ scalars.
 A third option,
whose consistency needs however to be checked in string theory,
 can a priori also be envisaged:
since $H$ does not break the reflection symmetry
$(X_4,X_5) \rightarrow (-X_4,-X_5)$ of the world-sheet
action, eq. (3.1), we could mod it out
and convert
  the torus  into a $Z_2$ orbifold.
This
projects out of the spectrum the non-zero helicity components of
all $6d$ gauge bosons, thus eliminating the dangerous, potentially
tachyonic states. Note that since the
 area of the orbifold is half that
  of the torus,
the minimal magnetic charge must in this case be doubled.
If despite all of the above cures there remain tachyonic scalars
in the spectrum, they will acquire non-zero expectation values
in the vacuum. In contrast to Wilson-line breaking, this mechanism
can reduce the rank of the gauge group as the example in the
following section will illustrate.

\setcounter{footnote}{0}

Setting magnetic instabilities aside for the moment, let me comment
briefly on another important issue, i.e. the cancellation of
anomalies.
Suppose there are no pure gauge anomalies in six
dimensions, so that the relevant box diagram is zero.
 Since the magnetic
field gives mass to all charged gauge bosons, the unbroken
gauge group after compactification
is necessarily of the form
$  U(1)_{H}\times G$. Let the $6d$ chiral fermions transform in the
representations
$\oplus (q_{({\cal R})},{\cal R})$ of this gauge group.
According to the index theorem, eq. (2.9),
the net number of
(left minus right) $4d$ chiral fermions
 in the  $(q_{\cal R},{\cal R})$
representation is   $q_{({\cal R})} K$.
The $G-G-G$ triangle anomaly
thus reads:
$$\eqalign{
(GGG)\  {\rm triangle} \ \ \propto
 \ K \sum_{\cal R}\ &q_{({\cal R})} \ tr_{\cal R} (T^{\{\alpha}
T^\beta T^{\gamma\}}) \cr
&\propto \ \ \Bigl( GGG-U(1)_H\Bigr) \ {\rm box} \ = 0 \ , \cr}
\eqno(4.1)
$$
where   the $T^\alpha$ are
  generators of $G$.
Likewise one can    show easily that all triangle
anomalies involving $U(1)_H$ vanish. This is
a special case of
the argument given for arbitary compactifications by Witten
\cite{O32,GSW}.
 The only
subtle point is the fact that anomaly
cancellation in ten dimensions makes use of the two-index
antisymmetric-tensor $B_{\mu\nu}$, with modified
field strength
$$ {\cal H}_{\mu\nu\rho} =
\partial_{[\mu} B_{\nu\rho ]} - tr \Bigl(
A_{[\mu} \partial_{\nu}A_{\rho ]} +
 {2\over 3} A_{[\mu}A_{\nu}A_{\rho ]}
\Bigr) \ , \eqno(4.2)$$
where we neglect here gravitational backgrounds.
Since ${\cal H}_{\mu\nu\rho}$ contributes to the energy it must
be globally well-defined, which means that  \cite{GSW,O32}
$$\int_{\cal K}\  d\wedge {\cal H} = \int_{\cal K}\
 tr(F\wedge F) = 0
\  \eqno(4.3)$$
  for any compact four-manifold ${\cal K}$. It follows that
consistent compactifications on several magnetized tori
 must obey
\footnote{We use $F_a$ to denote the Lie-algebra valued magentic
field, and $H_a$ its appropriately normalized magnitude.}
$$ tr (F_a F_b) = 0 \ \ {\rm for} \ a\not=b \ , \eqno(4.4)$$
i.e. the various magnetic fields should point in orthogonal
directions in group space.
This guarantees that in the $4d$ theory any residual anomalies can
still be cancelled by the mechanism of Green and Schwarz.

The reader may wonder why anomalies are treated differently from
  other ultraviolet divergences of the annulus or M\"obius-strip,
  which signal non-vanishing gravitational tadpoles.
The reason in
 ten dimensions
is that anomalies give rise to tadpoles of
unphysical Ramond-Ramond states  \cite{annulus},
 and cannot therefore,  even in principle, be cured
by shifting gravitational backgrounds.
In four dimensions, on the other hand, they are a signal of
an ill-defined compactification as just noted.
Another fact concerning eq. (4.3) is also worth pointing out:
let the four-manifold ${\cal K}$ be the product of
a torus with magnetic field $H$ on its surface,
 and of a sphere at spatial infinity.
If all $4d$ scalars go to constant values at
infinity, the integral factorizes into the product of magnetic
fluxes. In order to satisfy (4.3)
 the compactified theory should therefore have
no monopoles with magnetic charge under $U(1)_H$.
Does this mean that
electric-magnetic duality is
  necessarily broken in this case?
The answer is not obvious
because the term ${\cal H}^2$ in the
Lagrangian gives a mass of order
$H \sim m_{SUSY}^2/M_{Planck}\sim 10^{-3} eV$
to the $U(1)_H$ gauge boson, with  $B_{45}$  furnishing its
longitudinal component.
This is a tiny mass indeed, since
$B_{\mu\nu}$ has gravitational-strength
couplings, but  electric charges are all the same
screened and duality could be possibly  saved.

The last thing I would like to discuss in this section, is the
issue of radiative corrections.
These pose a threat to perturbative calculability, since above the
supersymmetry threshold the theory is higher-dimensional and hence
non-renormalizable, at least naively.
Let me concentrate in particular on maximal $N=4$ supersymmetry, as
this might offer the best hope of alleviating the problem.
 To keep the
expressions simple I suppose   that only one of the tori
is magnetized, say $F_{45}\not=0$,
  but the results will also hold in the more general
 situation.
All particles belong to   spin-1 multiplets of $N=4$,
which contain
 one gauge boson, four Weyl fermions and six scalars.
For massive multiplets one of the scalars is eaten by
the longitudinal component of the vector.
Out of the eight
bosonic states two have internal
helicity $\Sigma_{45}=\pm1$ and all
the others zero,
while   the eight fermionic states have
 $\Sigma_{45}=\pm{1\over 2}$ in equal numbers.
A simple counting then shows that at each Landau level separately
$$ str {\cal M}^{2n} = 0 \ \ \ {\rm for\ \  all } \ \ n<4 \ \ .
 \eqno(4.5)$$
As a result the one-loop ultraviolet
behaviour is indeed much more soft
than naively expected.
Consider for instance a multiplet of mass $M_0$ in six
dimensions.
Using   the expansion
$$ str \ e^{- z \Sigma_{45}} = \sum_{l=2}^\infty \ 2 (1-4^{1-l})\
 {z^{2l}\over (2l)!}
\ , \eqno(4.6)$$
 and the Schwinger proper-time parametrization,
 one finds the following result for its
contribution  to the
one-loop
vacuum energy:
$$\eqalign{
\Lambda_{1-loop} &=
\sum_{n=0}^\infty\  {1\over 32\pi^2}\
\int_0^\infty {dt\over t^3}\
e^{-t\Bigl(M_0^2+(2n+1)\vert qH\vert \Bigr)}\
str ( e^{-2t qH \Sigma_{45}} ) \cr
& = {1\over 16\pi^2}\
\sum_{l=2}^\infty \   (2^{2l}-4 )\
{(2l-3)! \over (2l)!} \vert qH \vert^{2l}\
\sum_{n=0}^\infty \ [ M_0^2 + (2n+1)\vert qH \vert ]^{2-2l} \ .
\cr} \eqno(4.7)
$$
Since the double summation is absolutely convergent, the
result is ultraviolet finite despite the infinite tower of Landau
levels. Furthermore for massive multiplets, $M_0\gg m_{SUSY}$,
the result vanishes like the sixth power of the supersymmetry scale:
$$
 \Lambda_{1-loop} \simeq {\vert qH\vert^3\over 64\pi^2 M_0^2}
 \sim m_{SUSY}^6/M_0^2 \ \ . \eqno(4.8)$$
Likewise one can show that the one-loop
contribution of each $6d$
multiplet to the effective   gauge coupling
is finite. A $6d$ multiplet in a representation ${\cal R}$
of some simple factor $G$ of the gauge
 group contributes indeed the
following correction to $\alpha_G^{-1}$:
 $$\eqalign{
\Delta^G_{1-loop} &=
{c_2({\cal R})\over 2\pi} \ \sum_{n=0}^\infty
 \int_0^\infty {dt\over t}\ e^{-t[M_0^2 + (2n+1)\vert qH\vert]}
\ str \Bigl[ ({1\over 12}- \chi^2) \ e^{-2tqH\Sigma_{45}} \Bigr]
\cr &=
 {c_2({\cal R})\over 12\pi}\
 \sum_{l=1}^\infty \   {(2^{2l}+8)  \over 2l } \vert qH \vert^{2l}\
\sum_{n=0}^\infty \ \Bigl[ M_0^2 +
 (2n+1)\vert qH \vert \Bigr]^{-2l} \ ,
\cr} \eqno(4.9)$$
where
$c_2({\cal R})$
is the quadratic Casimir of the representation ${\cal R}$ , and
  $\chi$ is here the   helicity in four dimensions
\footnote{Note that $\Delta$ can be obtained as
 the coefficient of the ${\cal B}^2 V/8\pi$
term in the one-loop vacuum energy, when one turns on a
magnetic field ${\cal B}$ in a three-dimensional volume V.  This can
be used to check the relative normalizations of eqs. (4.7) and (4.9).}.
The double summation
 is once more convergent,
and the result
  is finite even though we are dealing with
a $6d$ theory and supersymmetry is broken.
Furthermore  for
particles much
above the susy threshold we find
$$ \Delta_{1-loop}^G \simeq {c_2({\cal R})\over 8\pi}
 {\vert qH\vert\over M_0^2} \sim
   m_{SUSY}^2/M_0^2 \ \ , \eqno(4.10)$$
i.e. such particles do not even contribute  a finite
renormalization to the gauge couplings!

Does this mean that the   couplings will stay small as
we go up in energy towards the Planck scale?
Things are, unfortunately,  not so simple.
First, this is a one-loop result, and there is certainly no
guarantee that divergences will not show up at two- or
higher-loop order.
 Second, if more than two compact dimensions become
large, we must sum the above expressions over the
$6d$ masses ($M_0$) of new towers of Kaluza-Klein states.
The correction will thus grow logarithmically if there are two
extra large radii, and it would grow like the square of energy if
all six compact dimensions were large.
But six large compact dimensions is precisely
 what we need in order to reconcile
chirality with low-energy $N=4$ supersymmetry,
 as previously noted.
This example
serves in fact to illustrate a crucial point:
since for fixed torus area the magnetic field  is a discrete
modulus, {\it the limit of supersymmetry restauration cannot be taken
independently and need not commute with the sum over
 Kaluza-Klein states.}
Put differently, although every   multiplet makes a
vanishingly-small contribution to the running, the fact that
there can be as many as $(M_{Planck}/m_{SUSY})^6$ of them can
lead to a very large cummulative effect in the decompactification
limit.
The absence of renormalization in the
supersymmetric compactification
$H_a=0$, does not in particular
 protect us agaist potential disaster.
In chiral Scherk-Schwarz compactifications power-law corrections
can be suppressed at one loop \cite{A}, but this
difficulty should show up at two- and higher-loop order.

\vskip 0.6cm
{\bf 5. A standard model with broken N=4}

I will now describe a  compactification of the
$SO(32)$ superstring \footnote{See also ref.
\cite{O32} for an earlier
effort.},
 that exhibits the two main features
of magnetized tori: the reconciliation of chirality
with extended low-energy supersymmetry, and the triggering
of electroweak breaking by the Nielsen-Olesen instability.
I label the three magnetized tori
   by lower-case Latin indices, as in section 3,
and consider the minimal fields allowed
by the Dirac quantization condition:
 $$ F_{a} =    {\cal A}_{a}^{-1} Q_a \ ,
 \ \ \ {\rm for}\ \ a =(45),(67),(89) . \eqno(5.1)
$$
Here  ${\cal A}_a$ is the area of the $a$th torus, and
$Q_{a}$  is the corresponding generator  of $SO(32)$,
 normalized  so that the elementary charge
in the adjoint representation  equals one.
In order to define the $Q_a$ we will
   use the following sequence of
embeddings:
$$\eqalign{
SO&(32)  \supset SO(10) \times SO(6)_{hor} \times SO(8)
\times SO(8)^\prime
\cr &\supset
\Bigl[ SU(5) \times  U(1) \Bigr]
\times \Bigl[ SU(3) \times U(1) \Bigr]_{hor}
 \times  \Bigl[ SU(4) \times
 U(1)\Bigr]
\times  \Bigl[ SU(4) \times
 U(1)\Bigr]^\prime \cr}
 \    $$
I assume a maximal embedding
$SO(2N) \supset SU(N)\times U(1)$, and
will denote by
$T_{(5)}$, $T_{(3)}$, $T_{(4)}$ and $T_{(4)}^\prime$ the
four
  $U(1)$ generators
in the order in which they appear   above from left to right.
Note that when these are normalized to unit charge,
their trace in the adjoint of $SO(32)$ reads
$$tr_{adj}\ T_{(N)}^{\ \ 2}= 60N \ \ , \eqno(5.2)$$
where the $N$ refers to the subgroup $SO(2N)$ in which the
generator is
embedded.
Let us now choose the directions of the three magnetic fields as
follows:
$$\eqalign{
 Q_{(45)} &=
{1\over 2}  \Bigl( 3T_{(5)}-5T_{(3)}+T_{(4)}-T_{(4)}^\prime \Bigr)
 \cr
 Q_{(67)} & = {1\over 2}
 \Bigl( T_{(5)}+T_{(3)}-T_{(4)}-T_{(4)}^\prime \Bigr)\cr
 Q_{(89)} & =  {1\over 2}
 \Bigl(T_{(5)}+T_{(3)}+T_{(4)}+T_{(4)}^\prime
\Bigr)
  \ \ . \cr} \eqno(5.3)
$$
Using eq. (5.2) one can   check easily that these
 are orthogonal generators of $SO(32)$,
 so that the anomaly condition is satisfied.

Now recall that chiral fermions can
only come from multiplets whose charges $q_a$ under
all three magnetic $U(1)$'s do not vanish. This is
necessary in order to fix   all internal helicities of
the $10d$ Weyl spinor, and can be also seen from the
expression for the  net chirality in four dimensions:
$$n_L-n_R = q_{(45)} q_{(67)} q_{(89)} \ .
 \eqno(5.4)$$
As a result there are no chiral
 fermions transforming under both the
"observable" gauge group $[SU(5)\times U(1)]\times
[SU(3)\times U(1)]_{hor}$, and the
  "hidden"
one $\Bigl[ SU(4)\times U(1)\Bigr]^2$.
Simple inspection shows in fact that there are only three
types of chiral fermions in the "observable sector", whose charges
and multiplicities are listed in the table below:

\vskip1.2cm

\vbox{\tabskip=0pt
\offinterlineskip
\def\tablerule{\noalign{\hrule}}
\halign to420pt{\strut#&\vrule#\tabskip=1em plus2em&
\hfil#& \vrule#& \hfil#\hfil& \vrule#&
\hfil#& \vrule#&
\hfil#& \vrule#&
\hfil#& \vrule#\tabskip=0pt\cr
\tablerule
&&\omit&&\omit&&\omit&&\omit&&\omit&\cr
&&\omit\hidewidth $SU(5)\times SU(3)_{hor}$ \hidewidth&&
\omit\hidewidth $q_{(45)}$\hidewidth&&
\omit\hidewidth $q_{(67)}$ \hidewidth&&
\omit\hidewidth $q_{(89)}$\hidewidth&&
\omit\hidewidth {\# of states}\hidewidth&\cr
&&\omit&&\omit&&\omit&&\omit&&\omit&\cr
\tablerule
&&\omit&&\omit&&\omit&&\omit&&\omit&\cr
&&$(10,1)$&&3&&1&&1&&3&\cr
&&\omit&&\omit&&\omit&&\omit&&\omit&\cr
\tablerule
&&\omit&&\omit&&\omit&&\omit&&\omit&\cr
&&$({\bar 5},{\bar 3})$&&1&&-1&&-1&&1&\cr
&&\omit&&\omit&&\omit&&\omit&&\omit&\cr
\tablerule
&&\omit&&\omit&&\omit&&\omit&&\omit&\cr
&&$(1,3)$&&5&&-1&&-1&&5&\cr
&&\omit&&\omit&&\omit&&\omit&&\omit&\cr
\tablerule
  \noalign{\smallskip}
\cr}}

\vskip 0.3cm
{\bf Table 1.} Chiral fermions
transforming under the "observable" gauge group,
their charges under the three magnetic
 $U(1)$'s and their multiplicities.
\newpage

This is precisely the anomaly-free content of an $SU(5)$
grand-unified model
with three chiral families of quarks and leptons,
 and an extra horizontal
$SU(3)_{hor}$ symmetry under
 which transform   the
  ${\bar 5}$'s as well as standard-model singlets.
 To proceed further
we would like to break  $SU(5)\rightarrow SU(3)_c \times SU(2)_w\times
U(1)_Y$,
render massive other unwanted
 massless states and take care of   Nielsen-Olesen
instabilities.
To these ends we still have Wilson lines,
 and the choice of the radii
 at our disposal. Rather than being systematic, let me make
some choices and see where they lead us.
Denoting by $Y$
the hypercharge generator inside $SU(5)$, we set
$$ a_6 = a_8 =  b\ T_{(4)} + b^\prime T_{(4)}^\prime \ ,
\ \
 a_7 = -c\  T_{(5)} + c_Y  Y \ \
{\rm and} \ \ a_9\in SU(3)_{hor} \ \ , \eqno(5.5)$$
where $b,b^\prime,c$ and $c_Y$
are constants.
Recall that a Wilson line is relevant only when the
charge under the  magnetic field on the
 corresponding torus does not vanish.
Chiral fermions and all their $N=4$ partners are
 not therefore affected by the above
 backgrounds.
Gauge bosons of unbroken symmetries, on the other hand,
have all three $q_a=0$, so the $a_7$ and $a_9$ Wilson
lines will break $SU(5)$ to the standard model group, and the
horizontal symmetry to $U(1)$ factors. Furthermore
all particles  charged under both the hidden
 and observable gauge groups
have either $q_{(67)}=0$ or $q_{(89)}=0$,
 so they will obtain a mass
from either the $a_8$ or the $a_6$ Wilson line.
 By choosing moduli appropriately
we can ensure that none of these states is tachyonic.

What about the breaking of electroweak symmetry?
Candidate higgs doublets  come from two different places:
  scalar partners of the chiral fermions in the representation
$({\bar 5},{\bar 3})$ of $SU(5)\times SU(3)_{hor}$, and
 scalars in the $({\bar 5},3)$ multiplet,
which does not contain chiral fermions since it has
  $q_{(45)} = -4$ and
$q_{(67)}=
q_{(89)}=0$.
The  $({\bar 5},3)$ scalars
 are  preferable for two reasons: {\it (i)} they
have non-vanishing (renormalizable)
Yukawa couplings with the chiral fermions as can be seen from
their $SU(3)_{hor}$ transformation properties,
and {\it (ii)} the
Wilson line $a_7$   splits the colour-triplet from the doublet,
and can thus ensure that $SU(3)_c$ remains unbroken.
Using the formulae (3.11) and (2.2),
 and the fact that $Y=-1$
or ${2\over 3}$ for the doublet or
(anti)triplet in the ${\bar 5}$ of $SU(5)$,
we find the following masses for the lowest-lying scalars
that transform non-trivially under the standard model gauge group:
 $$\eqalign{
 {\cal M}^2_{({\bar 5},{\bar 3})} &=
  {\cal A}_{(45)}^{-1} + {\cal A}_{(67)}^{-1}
+ {\cal A}_{(89)}^{-1} - 2\  {\rm max}
\Bigl( {\cal A}_{a}^{-1} \Bigr)  \cr
{\cal M}^2_{(10,1)} &=
  3{\cal A}_{(45)}^{-1} + {\cal A}_{(67)}^{-1}
+ {\cal A}_{(89)}^{-1} - 2\  {\rm max}
\Bigl( 3{\cal A}_{(45)}^{-1}, {\cal A}_{(67)}^{-1},
{\cal A}_{(89)}^{-1}
 \Bigr) \ ,  \cr}
  \eqno(5.6)$$
 and
 $$ {\cal M}^2_{({\bar 5},3)} =
 \cases{
  &  $- 4 {\cal A}_{(45)}^{-1} + (c- c_Y )^2$  \ , \
\ {\rm (doublet)} \cr
   & $ - 4 {\cal A}_{(45)}^{-1} + ( c + {2\over 3}c_Y)^2$
\ ,\ \ {\rm (triplet)} \ \  \cr}
\eqno(5.7)$$
Note that in what concerns eq. (5.7),
 the contribution of the $a_7$
Wilson line cannot exceed $(\pi/R_7)^2$,   the contribution of
the $a_9$ background was omitted for simplicity, and   there
are $\vert q_{(45)}\vert = 4$ excitations in the lowest Landau level.
Now if we want only colour-singlet scalars to be tachyonic, we
must demand that both
$({\cal A}_{(45)}^{-1}, {\cal A}_{(67)}^{-1},
{\cal A}_{(89)}^{-1})$
and
$(3{\cal A}_{(45)}^{-1}, {\cal A}_{(67)}^{-1},
 {\cal A}_{(89)}^{-1})$
satisfy the triangle inequalities,  and that
$$ \vert c- c_Y \vert < 2/\sqrt{{\cal A}_{(45)}} <
\vert c + {2\over 3} c_Y \vert  \ . \eqno(5.8)$$
These constraints are   mutually compatible,
choosing for instance
${\cal A}_{(67)}={\cal A}_{(89)}<{2\over 3} {\cal A}_{(45)}$
and $c=c_Y= R_7^{-1} = R_6^{-1}$ will satisfy them all.
In an appropriate region of parameter space the higgs
doublets will thus acquire non-zero vevs,  breaking
electroweak symmetry and giving mass to the quarks and leptons
\footnote{By going to the corner of parameter
 space where the doublets
are just barely tachyonic, we can be sure that their non-zero vevs
will not have a big effect on the masses of all other scalars.}.
Furthermore, unless we fine tune parameters,
 the scale of electroweak breaking
$m_Z$ will be tied classically to $m_{SUSY}$.

Although this model has too much structure at the supersymmetry
threshold to be considered at this point
 as realistic, it is
surprising how close we come to a classical description of our
low-energy world with relatively little effort.
Several features which have proven hard to achieve
 in previous string
model-building, come out rather easily here:
adjoint scalars for $SU(5)$ breaking,
 three chiral families of quarks and leptons, and
negative mass square for the higgs doublets.
More conservative uses of magnetized tori can also be envisaged:
since for example mass splittings are proportional to charge,
one can restrict tree-level supersymmetry breaking to a hidden
and possibly hypermassive sector. Alternatively, one may insist
that there are only two large compact dimensions \cite{Fayet},
 so that the low-energy world is
$N=2$ supersymmetric.
Finally, it could be desirable to have both magnetic
fields and modified, Scherk-Schwarz boundary conditions:
 the former
would create chirality and trigger electroweak breaking, while
the latter would give tree-level mass to all
the standard-model gauginos.

\vskip0.6cm
{\bf 6. Outlook}

No proposal for breaking supersymmetry   avoids at present
all gravitational tadpoles,  for the dilaton, moduli and conformal factor
of the metric.
Allowing such tadpoles classically is thus
  a valid alternative,  and may be necessary for getting a
good approximation of our low-energy world from the string.
This logic opens up a host of possibilities of which perhaps the
simplest one, compactification of $SO(32)$ string theory
on magnetized tori, was studied in
  this paper in detail.
Two  remarkable features of such compactifications,
namely the reconciliation of chirality with extended low-energy
supersymmetry, and the Nielsen-Olesen-triggered
 electroweak breaking,
should survive in more general settings.
It should, in particular, be possible to extend
the results of this paper
to compactifications of the heterotic string on products of
magnetized two-spheres.
As
 in the limiting field theory \cite{horv}, the spectrum
  in the full string theory should also be calculable exactly,
  because
string motion on a magnetized sphere
can be described by a WZW model \cite{RW,Monopole}.
The heterotic embedding can, however, be subtle.
Whether the quantum string dynamics can stabilize such a vacuum,
without going to negatively-curved supersymmetric space-times
\cite{electrovac} is of course the big and open question.
The other major difficulty, shared by the
Scherk-Schwarz scenario \cite{ABLT,A}
and other marginal deformations of
classical string vacua \cite{Banks},
 comes from the fact that
$m_{SUSY}$ is tied to the size of compact dimensions.
In the compactification of section 5, for example, the entire
$10d$ $SO(32)$ string lies just beyond the supersymmetry threshold!
Needless to say this would have dramatic consequences, such as
 infinite towers of mirror fermions shifted relative to
each other by $m_{SUSY}$.
Unfortunately,  it also poses
  a threat to perturbative calculability,
 which as I explained above
cannot be addressed by studying the exact supersymmetric limit.

\vskip 0.5cm
{\bf Aknowledgements}

I have benefited from discussions with I. Antoniadis,
 P. Fayet, I. Pavel and
 A. Zaffaroni. This research was supported in part by
EEC grants SC1-CT92-0792 and CHRX-CT93-0340.

\end{document}